\documentclass[journal]{IEEEtran}
\usepackage{graphicx}
\usepackage[cmex10]{amsmath}
\usepackage{cases}
\usepackage{amsthm}
\usepackage{amssymb}
\usepackage{algorithm}
\usepackage{algorithmic}
\usepackage{multirow}
\usepackage{amsmath}
\usepackage{xcolor}
\usepackage{subeqnarray}
\usepackage{cases}
\usepackage{enumerate}
\usepackage{cuted}
\usepackage{booktabs}
\usepackage{mathrsfs}
\usepackage{bbm}
\usepackage{graphicx}
\usepackage{lettrine}
\usepackage[colorlinks]{hyperref}

\ifCLASSINFOpdf
\else
\fi

\begin{document}
\title{\LARGE Fluid Reconfigurable Intelligent Surfaces:\\Joint On-Off Selection and Beamforming with Discrete Phase Shifts}

\author{Han Xiao,~\IEEEmembership{Student Member,~IEEE,}
            Xiaoyan Hu,~\IEEEmembership{Member,~IEEE,}
            Kai-Kit~Wong,~\IEEEmembership{Fellow,~IEEE,}\\
            Hanjiang Hong,~\IEEEmembership{Member,~IEEE,}
            George C. Alexandropoulos,~\IEEEmembership{Senior~Member,~IEEE}, and\\
            Chan-Byoung Chae,~\IEEEmembership{Fellow,~IEEE}
\vspace{-7mm}

\thanks{The work of H. Xiao was supported by the China Scholarship Council (CSC) Scholarship under Grant 202406280315. The work of X. Hu was supported in part by the National Natural Science Foundation of China (NSFC) under Grants 62471380 and 62201449, 
the open research fund of National Mobile Communications Research Laboratory, Southeast University under Grant No. 2025D08. The work of K.-K. Wong was supported by the Engineering and Physical Sciences Research Council (EPSRC) under Grant EP/W026813/1. The work of C.-B. Chae was supported by IITP under Grants IITP-2025-RS-2024-00428780 and IITP-2021-0-00486 funded by the Korean government. \emph{(Corresponding author: Xiaoyan Hu.)}}
\thanks{H. Xiao and X. Hu are with the School of Information and Communications Engineering, Xi'an Jiaotong University, Xi'an 710049, China. X. Hu is also with the National Mobile
Communications Research Laboratory, Southeast University, Nanjing 210096, China (e-mails: $\{\rm hanxiaonuli@stu.xjtu.edu.cn; xiaoyanhu@xjtu.edu.cn\}$).}
\thanks{ K. K. Wong and H. Hong are with the Department of Electronic and Electrical Engineering, University College London, London, United Kingdom. K.-K. Wong is also with Yonsei Frontier Lab, Yonsei University, Seoul, Korea (e-mail: $\{\rm hanjiang.hong, kai\text{-}kit.wong\}@ucl.ac.uk$).}
\thanks{G. C. Alexandropoulos is with the Department of Informatics and Telecommunications, National and Kapodistrian University of Athens, 16122 Athens, Greece and the Department of Electrical and Computer Engineering, University of Illinois Chicago, IL 60601, USA (e-mail: $\rm alexandg@di.uoa.gr$).}
\thanks{C.-B. Chae is with the School of Integrated Technology, Yonsei University, Seoul, 03722, Korea (e-mail: $\rm cbchae@yonsei.ac.kr$).}
}

\maketitle

\begin{abstract}
This letter proposes a fluid reconfigurable intelligent surface (FRIS) paradigm, extending the conventional reconfigurable intelligent surface (RIS) technology to incorporate position reconfigurability of the elements. In our model, a `fluid' element is realized by a dense matrix of subelements over a given space and dynamically selecting specific elements for signal modulation based on channel conditions. Specifically, we consider a FRIS-assisted single-user single-input single-output (SU-SISO) system and formulate an optimization problem that can jointly optimize element selection and their discrete phase shifts to maximize the achievable rate. To address this problem efficiently, we propose an iterative algorithm based on the cross-entropy optimization (CEO) framework. Simulation results reveal that FRIS achieves significant performance gains over traditional RIS.
\end{abstract}

\begin{IEEEkeywords}
Fluid antenna system (FAS), fluid reconfigurable intelligent surface (FRIS), position optimization, discrete phase-shift, cross-entropy optimization (CEO).
\end{IEEEkeywords}

\IEEEpeerreviewmaketitle
\vspace{-2mm}
\section{Introduction}\label{sec:S1}
\IEEEPARstart{T}{hough} reconfigurable intelligent surface (RIS) technology offers a cost-effective and energy-efficient solution to enhance wireless coverage and capacity, practical applications still face significant challenges. The first major challenge is the complexity of channel estimation between RISs and other terminals, such as users and base stations (BSs)~\cite{wei2021channel}. Since RIS is inherently passive, achieving satisfactory performance often requires deploying a large number of reflecting elements. However, increasing element density significantly raises pilot overhead and overall system complexity.

Another critical challenge associated with RIS is the `multiplicative' or `doubly' fading effect, where the total path loss of the transmitter-RIS-receiver link equals the product (rather than the sum) of the individual path losses of the transmitter-RIS and RIS-receiver links. This multiplicative effect often results in severe signal attenuation~\cite{zhang2022active}. A promising strategy to overcome these challenges involves introducing additional degrees of freedom (DoFs) into RIS-based systems, enhancing performance without significant hardware complexity \cite{Basar-2024}.

In this context, the concept of fluid antenna system (FAS) appears promising. FAS represents a new generation of reconfigurable antennas characterized by shape and position flexibility, introducing additional DoFs at the physical layer~\cite{wong2020fluid,New2024aTutorial}. An electromagnetic theoretical interpretation of FAS was provided in \cite{Lu-2025}. In recent years, considerable progress has been made in understanding the potential of FAS. For instance, the diversity-multiplexing trade-off of FAS was explored in~\cite{new2023information}, while channel coding combined with FAS for massive multiple access was investigated in~\cite{hongcoded2025}. Moreover, FAS has been shown to be effective in multi-carrier systems~\cite{Hong-2025arxiv}  and for channel estimation as well~\cite{xu2024channel, New-twc2025}. Other recent applications include mobile edge computing~\cite{fluidzuo2024}, integrated sensing and communication (ISAC)~\cite{Zou-2024}, and physical layer security \cite{Tang-2023}. %

Encouraged by the position reconfigurability of FAS, Ye \textit{et~al.}~\cite{ye2025joint} introduced the concept of fluid RIS (FRIS) and applied it to an ISAC system. This initial approach proposed physically movable RIS elements capable of adaptively adjusting the positions of reflecting elements. But from a practical implementation viewpoint, physically moving the elements raises real concerns. The hardware complexity, slow response times, precision difficulties, increased power consumption, and mechanical reliability and maintenance costs all cast doubt on the practicality of such systems. Subsequently, Salem \textit{et al.}~\cite{salem2025first} revisited the fluid RIS concept, redefining fluid elements as sub-areas of the RIS surface where reflection points can be dynamically switched amongst many closely spaced ports without physical movement. Their results demonstrated substantial performance gains compared to conventional RIS, suggesting a positive outlook for FRIS technology. Nevertheless, the idealized assumption of perfect phase shifters used in~\cite{salem2025first} raises concerns regarding practical feasibility.


Motivated by these considerations, in this letter, we propose a novel FRIS framework, wherein the reflecting elements achieve position reconfigurability through numerous densely distributed ports within the available RIS surface. To achieve superior spatial flexibility, the spacing between adjacent ports is deliberately designed to be significantly smaller than half a wavelength, creating a quasi-continuous RIS surface--a notable distinction from existing FRIS implementations. Furthermore, a unique aspect of our framework is that each activated fluid element incorporates a discrete phase shifter, accurately representing practical hardware constraints. The main contributions of this letter can be summarized as follows:
\begin{itemize}
\item We present the signal transmission model for the proposed FRIS-assisted communication system.
\item We formulate an optimization problem for maximizing the achievable rate, with practical constraints.
\item We then propose a novel iterative algorithm based on the cross-entropy optimization (CEO) framework to jointly optimize element positions and discrete phase shifts.
\item Finally, we demonstrate the effectiveness and great potential of the proposed FRIS through performance comparisons against a conventional RIS.
\end{itemize}

\section{System Model}\label{sec:S2}
\begin{figure}[!t]
\centering
\includegraphics[scale=0.25]{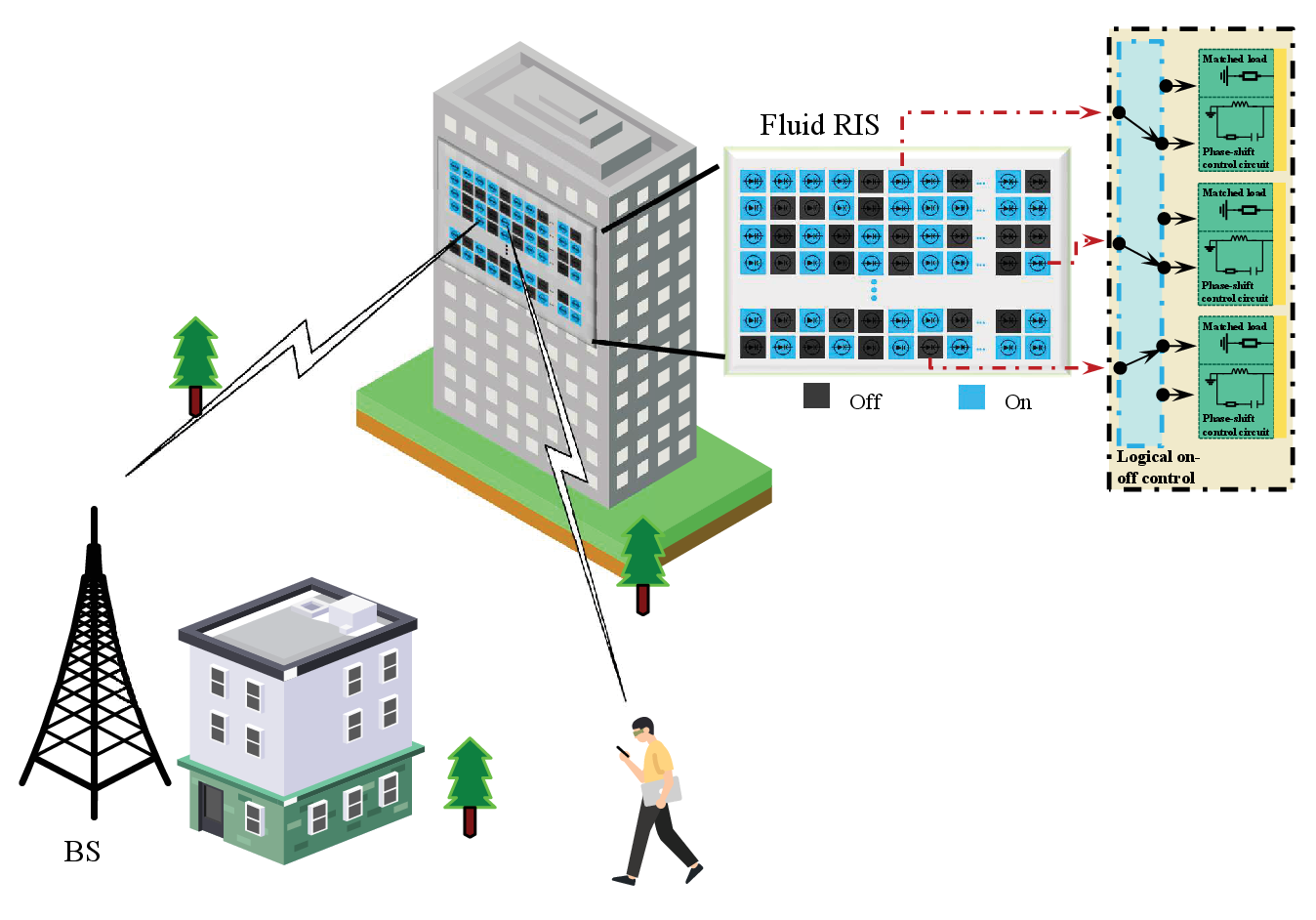}
\caption{The considered model of a FRIS-assisted wireless network.}\label{fig:scenario}
\end{figure}

The system model is depicted in Fig.~\ref{fig:scenario}, which comprises a BS with a fixed-position antenna (FPA), a single-FPA user, and a FRIS. The FRIS has $M=M_y\times M_z$ on-off discrete-response reflecting elements that are uniformly distributed over the surface. This is a classical single-user single-input single-output (SU-SISO) system with the aid of a FRIS. The inter-element spacing is $d$ which is less than $\frac{\lambda}{2}$ where $\lambda$ denotes the carrier wavelength. It is worth noting that each element in the FRIS is designed with two operational modes: `off' and `on'. In the `off' mode, the unit is connected to a matched load circuit, and it does not modulate the incident signal. Conversely, in the `on' mode, the element will actively modify the electromagnetic properties of the incident signals, optimizing performance. Compared to \cite{salem2025first}, an element in our model may be viewed as a port in each fluid element. In this letter, the direct link between the BS and user is assumed completely broken by obstacles like buildings.

With the extremely small spatial distance between adjacent elements, $d$, the spatial correlation between any two elements cannot be ignored. Here, the Jake's model as a result of rich scattering \cite{new2023information} is adopted to characterize the spatial correlation between any two elements. Specifically, the spatial correlation of the $i$-th element and the $j$-th element is given by
\begin{align}
J_{i, j} = j_0\left(2\pi d_{i, j}/\lambda\right),
\end{align}
with
\begin{multline}
d_{i, j}^2=d^2\left(\operatorname{mod}(i, M_y)-\operatorname{mod}(j, M_z)\right)^2+\\
d^2\left(\left\lfloor\frac{i}{M_y}\right\rfloor-\left\lfloor\frac{j}{M_z}\right\rfloor\right)^2,
\end{multline}
 where $j_0(\cdot)$ denotes the zero-order spherical Bessel function of the first kind, $\operatorname{mod}(\cdot, \cdot)$ stands for complementary operator, $\lfloor\cdot\rfloor$ is the floor operator, which rounds up to the nearest integer. Thus, the spatial correlation matrix is given by
 \begin{align}
\mathbf{J}=\begin{bmatrix}
j_{11} & j_{12} &\cdots &j_{1M} \\
j_{21} & j_{22} &  \cdots & j_{2M} \\
\vdots & \vdots  & \ddots & \vdots\\
j_{M1} & j_{M2} & \cdots & j_{MM}
\end{bmatrix}.
 \end{align}

The channels between the FRIS and the BS or user without spatial correlation, i.e., $\mathbf{h}_\mathrm{br}$ and $\mathbf{h}_\mathrm{ru}$, are molded as
\begin{align}
\mathbf{h}_{\xi}=\sqrt{l_{\xi}}\mathbf{g}_{\xi}\in\mathbb{C}^{M\times 1},~ {\xi}\in\{\mathrm{br}, \mathrm{ru}\},
\end{align}
in which $l_{\xi}=\sqrt{\frac{\rho}{d_{\xi}^{\alpha}}}$ denotes the large-scale path loss with $\rho$ being the reference gain at one meter, $\alpha$ being the path-loss exponent, $d_{\xi}$ being the spatial distance between the FRIS and the BS/user, and $\mathbf{g}_{\xi}\in\mathbb{C}^{M\times 1}$ is the complex fading gain vector with each entry following a circularly symmetric complex Gaussian distribution with zero mean and unit variance.

It is assumed that $\widehat{M}$ elements will be turned on to modulate the incident signal. Considering the spatial correlation among elements, the received signal at the user can be written as
\begin{align}
y = \sqrt{P}\mathbf{h}_\mathrm{ru}^H(\mathbf{J}^\frac{1}{2})^H\mathbf{E}_{\widehat{M}}^T\mathbf{U}\mathbf{E}_{\widehat{M}}\mathbf{h}_\mathrm{br}s + n,
\end{align}
where $\mathbf{U} ={\rm Diag}\{e^{j\phi_1}, \dots, e^{j\phi_{\widehat{M}}}\} \in\mathbb{C}^{\widehat{M}\times\widehat{M}}$ is the reflected coefficient matrix, $\mathbf{E}_{\widehat{M}}^T=[\mathbf{e}_1, \dots, \mathbf{e}_{\widehat{m}}, \dots, \mathbf{e}_{\widehat{M}}]\in\mathbb{R}^{M\times\widehat{M}}$ represents the element selection matrix, where $\mathbf{e}_{\widehat{m}}, \widehat{m}\in\mathcal{\widehat{M}}=\{1, \dots, \widehat{M}\}$ is one of the columns of the $M\times M$ identity matrix. Note that any two columns of $\mathbf{E}_{\widehat{M}}^T$ must be different to prevent repeated selection of the same locations. In addition, $P$ represents the power budget at the BS, $n\sim \mathcal{CN}(0, \sigma^2)$ is the additional white Gaussian noise (AWGN) with $\sigma^2$ being the noise power. Utilizing singular value decomposition, $\mathbf{J}^\frac{1}{2}$ can be rewritten as $\mathbf{J}^\frac{1}{2}=\sqrt{{\bf\Lambda}}\mathbf{V}$, where matrix ${\bf\Lambda}$ is a diagonal matrix that consists of the eigenvalues, $\mathbf{V}\in\mathbb{C}^{M\times M}$ is a unitary matrix containing the eigenvectors of matrix $\mathbf{J}$.

Hence, the achievable rate of the user can be found as
\begin{align}
R=\log_2\left(1+P\left|\mathbf{h}_\mathrm{ru}^H(\mathbf{J}^\frac{1}{2})^H\mathbf{E}_{\widehat{M}}^T\mathbf{U}\mathbf{E}_{\widehat{M}}\mathbf{h}_\mathrm{br}\right|^2/\sigma^2\right).
\end{align}

\section{Problem Formulation and Algorithm Design}
\subsection{Problem Formulation}
To understand the benefits of FRIS, an optimization problem is formulated to maximize the achievable rate, considering the position selection of FRIS elements and their discrete phase-shift constraints. This problem can be expressed as
\begin{subequations}\label{eq_orig_opti}
\begin{align}
\max _{\mathbf{U}, \mathbf{E}_{\widehat{M}}} &~ R\label{eq_obj}\\
\text {s.t.} & \left|\mathbf{U}_{\widehat{m}, \widehat{m}}\right|=1, \forall \widehat{m}\in\mathcal{\widehat{M}},  \label{eq_orig_opti_1}\\
& \phi_{\widehat{m}}\in\mathcal{P}=\left\{v\frac{2\pi}{2^{b}}|v\in\left\{1, 2, \dots, 2^b\right\}\right\}, \forall \widehat{m}\in\mathcal{\widehat{M}},\label{eq_orig_opti_2}\\
& \mathbf{E}^T_{\widehat{M}}(:, l)\in\{\mathbf{e}_1, \dots, \mathbf{e}_{m}, \dots, \mathbf{e}_{M}\}, l\in\mathcal{\widehat{M}}, \label{eq_orig_opti_3}\\
& \mathbf{E}^T_{\widehat{M}}(:, l)\neq\mathbf{E}^T_{\widehat{M}}(:, g), \forall l\neq g,\label{eq_orig_opti_4}
\end{align}
\end{subequations}
where $b$ denotes the number of bits specifying the phase-shift resolution. Constraint \eqref{eq_orig_opti_4} is leveraged to enforce that the selected elements are different. Solving \eqref{eq_orig_opti} poses a considerable challenge when utilizing traditional existing optimization algorithms, see e.g., \cite{xiaoenergy2025} and references therein. This difficulty stems from the non-convex nature of the objective function (\ref{eq_obj}), the presence of the constant modulus constraint in \eqref{eq_orig_opti_1}, the involvement of discrete phase-shift variables, element selection considerations, and the strong interdependence between the variables. To overcome this complex optimization, an effective algorithm based on the CEO framework is proposed to optimize all the variables without resorting to alternative strategies. The details are presented in the next section.

\subsection{Algorithm Design}
The CEO framework is a probabilistic learning method with several key steps explained as follows \cite{xiaofrequency2024}:
\begin{itemize}
\item[(i)] \textbf{Sampling Distribution Design:} First, construct an appropriate sampling distribution model characterized by tilting parameters based on the characteristics of the variables.
\item[(ii)] \textbf{Generating Potential Solutions:} Evaluate candidate solutions generated from the established probability distribution model by their objective function values.
\item[(iii)] \textbf{Updating the Distribution:} Adjust the tilting parameters through the minimization of cross-entropy between the existing distribution and a reference distribution obtained from the best-performing solutions.
\item[(iv)] \textbf{Iterative Sampling:} Generate new samples utilizing the updated parameters iteratively until the difference in the optimal objective function values between consecutive iterations reaches a predetermined threshold.
\end{itemize}

\subsubsection{Sampling distribution model for reflection coefficients} Here, we focus on establishing the sampling distribution model of the discrete phase shifts of reflection coefficients. The constant modulus constraints of reflection coefficients can be easily satisfied by setting each reflection coefficient as $e^{j\phi_{\widehat{m}}},~ \widehat{m}\in\mathcal{\widehat{M}}$. Specifically, let $\boldsymbol{\phi}=[\phi_1, \dots, \phi_{\widehat{m}}, \dots, \phi_{\widehat{M}}]^T$. Note that each element in vector $\boldsymbol{\phi}$ is discrete and must choose a value from the set $\mathcal{P}=\left\{\varphi_v=v\frac{2\pi}{2^{b}}|v\in\{1, 2, \dots, V=2^b\}\right\}$. To denote the probability of the $\widehat{m}$-th entry selecting each value in $\mathcal{P}$, we define a probability set $\{P_{\widehat{m}, 1}, \dots, P_{\widehat{m}, v}, \dots, P_{\widehat{m}, V} \}$, where $P_{\widehat{m}, v}=\operatorname{Pr}\left(\mathcal{I}_{\widehat{m}}(\boldsymbol{\phi})=\varphi_v\right)$ denotes the probability that the $\widehat{m}$-th entry is chosen as $v\frac{2\pi}{2^{b}}$. Additionally, these probability values must satisfy the constraint $\sum_{v=1}^{V}P_{\widehat{m}, v}=1$. Since the sampling process of each element in $\boldsymbol{\phi}$ is independent, the joint sampling distribution of generating $\boldsymbol{\phi}$ can be found as
\begin{align}
\mathscr{H}\left(\boldsymbol{\phi}, \mathbf{P}\right)=\prod_{\widehat{m}}^{\widehat{M}}\sum_{v=1}^{V}P_{\widehat{m}, v}\mathbbm{1}_{\{\mathcal{I}_{\widehat{m}}(\boldsymbol{\phi})=\varphi_v\}},
\end{align}
where $\mathbf{P}$ is a matrix composed of $P_{\widehat{m}, v}, ~\widehat{m}\in\mathcal{\widehat{M}}, ~v\in\mathcal{V}=\mathcal\{1, \dots, V\}$, $\mathbbm{1}_{\{\cdot\}}$ is the indicator function for an event, and $\mathcal{I}_{\widehat{m}}(\boldsymbol{\phi})$ denotes the $\widehat{m}$-th entry of the vector $\boldsymbol{\phi}$.

Next, we will present the process of generating the discrete phase-shift for the $\widehat{m}$-th element based on its sampling distribution. Specifically, we begin by sampling a random number $z$ from a uniform distribution over the interval $[0,1]$. Then, leveraging this sampled value, we determine the discrete phase shift of the $\widehat{m}$-th element according to the following rule:
\begin{align}\label{eq_sampling_rule}
	\phi_{\widehat{m}} =\begin{cases}
		\frac{2\pi}{2^{b}},& z\leq P_{\widehat{m},1},\\
		2\frac{2\pi}{2^{b}},& P_{\widehat{m}, 1}\leq z\leq P_{\widehat{m}, 1}+P_{\widehat{m}, 2},\\
		~~\vdots & ~~~\vdots\\	
		V\frac{2\pi}{2^{b}},& \sum_{v=1}^{V-1}P_{\widehat{m}, v}\leq z\leq \sum_{v=1}^{V}P_{\widehat{m}, v}=1.
	\end{cases}
\end{align}
Based on this above rule, it can be observed that the probability for the random number, $z$, to fall within the interval $[\sum_{v=1}^{n-1}P_{\widehat{m}, v}, \sum_{v=1}^{n}P_{\widehat{m}, v}]$ is equal to that of $\phi_{\widehat{m}} =n\frac{2\pi}{2^{\lambda}}$. In other words, we have
\begin{align}
\operatorname{Pr}\left\{\phi_{\widehat{m}}=n\frac{2\pi}{2^{b}}\right\} &\hspace{.5mm}=\operatorname{Pr}\left\{\sum_{v=1}^{n-1}P_{\widehat{m}, v}\leq a\leq \sum_{q=1}^{n}P_{pq}\right\}\notag\\
&\stackrel{(a)}{=}\sum_{v=1}^{n}P_{\widehat{m}, v}-\sum_{v=1}^{n-1}P_{\widehat{m}, v}= P_{\widehat{m}, n},
\end{align}
where $(a)$ is due to the equal probability distribution of the uniform distribution. The outcome affirms that when implementing the generation rule outlined in \eqref{eq_sampling_rule}, the resulting $\phi_{\widehat{m}}$ adheres to its designated sampling distribution. By independently applying this method to individual elements, the system as a whole attains the targeted phase-shift arrangement in accordance with the joint sampling distribution.	

\subsubsection{Sampling distribution model for element selection} Let $\boldsymbol{\xi}\in\mathbb{R}^{M\times 1}$ be a vector associated with the element selection, where each entry is a binary variable that  chooses a value from the set $\{0, 1\}$. The value of the $m$-th entry of $\boldsymbol{\xi}$ indicates the $m$-th element's state of either `on' or `off'. When $\mathcal{I}_m(\boldsymbol{\xi})=1$, the $m$-th element is chosen to reflect signals, whereas  $\mathcal{I}_m(\boldsymbol{\xi})=0$ signifies that this element is in an inactive state. Actually, the element selection matrix $\mathbf{E}_{\widehat{M}}$ can be efficiently determined according to the value of $\boldsymbol{\xi}$. Considering that each component in $\boldsymbol{\xi}$ is a binary variable, the following sampling is utilized to generate $\boldsymbol{\xi}$, which is given by
\begin{align}
\mathscr{X}\left(\boldsymbol{\xi}, \mathbf{g}\right)=\prod_{m=1}^{M}g_{m}^{\mathcal{I}_m(\boldsymbol{\xi})}(1-g_{m})^{1-\mathcal{I}_m(\boldsymbol{\xi})},
\end{align}
where $\mathbf{g}=[g_1, \dots, g_m, \dots, g_M]^T$ with $g_{m}=\operatorname{Pr}\{\mathcal{I}_m(\boldsymbol{\xi})=1\}$ being the probability that the $m$-th entry in $\boldsymbol{\xi}$ equals $1$. We can utilize the similar approach of creating discrete phase shifts based on the sampling distribution to sample the varible $\boldsymbol{\xi}$.

Let $O$ represent the number of components in vector $\boldsymbol{\xi}$ that has a value of $1$, and it must adhere to the condition $O=\widehat{M}$. In fact, we cannot guarantee that every sampling solution adheres to this constraint. To obtain a feasible $\boldsymbol{\xi}$, we perform the following operations: (\romannumeral 1) If the sampled variable $\boldsymbol{\xi}$ satisfies the condition $O>\widehat{M}$, we will arrange the components of $O$ with a value of $1$ in decreasing order of probability and update the values of the entries to $0$ starting from the entries with the lowest probability until the condition $O=\widehat{M}$ is achieved. (\romannumeral 2) If $O$ is less than $\widehat{M}$, the elements of $M-O$ will be organized with a probability value of $0$ in ascending order, and the entries will be modified to $1$ beginning with the entries having the highest probability, until $O=\widehat{M}$ is met.

\subsubsection{Joint sampling distribution model} Consequently, we can obtain a joint sampling distribution to generate the feasible solution of the optimization problem \eqref{eq_orig_opti}, given by
\begin{align}\label{eq_joint_sd}
\mathscr{S}\left(\boldsymbol{\Gamma},\boldsymbol{\Xi}\right)=\mathscr{H}\times\mathscr{X},
\end{align}
where $\boldsymbol{\Gamma}\triangleq\{\boldsymbol{\phi}, \boldsymbol{\xi}\}$, and $\boldsymbol{\Xi}\triangleq\{\mathbf{P}, \mathbf{g}\}$ represents the tilting parameter of this joint sampling distribution.

\subsubsection{Updating formulas for tilting parameters} In this section, the sampling distribution in \eqref{eq_joint_sd} is utilized to produce a range of feasible solutions and then selecting these solutions with high performance to update the tilting parameters by minimizing the cross entropy between the current distribution and reference distribution. Specifically, we begin by sampling $A$ feasible solutions $\{\boldsymbol{\Gamma}_a\}_{a=1}^A$ from the joint distribution and assessing their respective objective function values. Based on these values, we arrange the $A$ feasible solutions in descending order as $\{\boldsymbol{\Gamma}_{[a]}\}_{a=1}^A$, resulting in $R(\boldsymbol{\Gamma}_{[1]})\geq R(\boldsymbol{\Gamma}_{[2]})\geq \cdots \geq R(\boldsymbol{\Gamma}_{[A]})$. Subsequently, the top $A_{\mathrm{elite}}=\zeta A$ solutions are selected to compose an elite set $\mathscr{E}=\{\boldsymbol{\Gamma}_{[a]}\}_{a=1}^{A_{\mathrm{elite}}}$, in which $\zeta\in(0, 1)$ represents the percentage of the highest-performing samples included in the elite set. Note that the elite set will be utilized to form the reference distribution. After that, we can fine-tune the tilting parameters through minimizing the cross-entropy between the existing distribution and the established reference distribution. According to \cite{rubinstein2004cross}, the cross-entropy minimization problem can be equivalently transformed as
\begin{subequations}\label{eq_entropy}
\begin{align}
\max _{\boldsymbol{\Xi}} &~\frac{1}{A}\sum_{a=1}^{A_\mathrm{elite}}\ln\mathscr{S}\left(\boldsymbol{\Gamma}_{[a]},\boldsymbol{\Xi}\right),\notag \\
\text {s.t. } &~\sum_{v=1}^{V}P_{\widehat{m}, v}=1,~ \forall\widehat{m}\in\mathcal{\widehat{M}}. \label{eq_entropy_1}
\end{align}
\end{subequations}
The optimization problem \eqref{eq_entropy} is a convex problem, since the objective function is a concave function with respect to (w.r.t.) variables $\boldsymbol{\phi}$ and $\boldsymbol{\xi}$. Therefore, the Lagrange multiplier method is adopted to address this problem. Specifically, the Lagrange function associated with \eqref{eq_entropy} can be expressed as
\begin{align}
\mathcal{L}=-\frac{1}{A}\sum_{a=1}^{A_\mathrm{elite}}\ln\mathscr{S}\left(\boldsymbol{\Gamma}_{[a]},\boldsymbol{\Xi}\right)+\sum_{\widehat{m}}^{\widehat{M}}\kappa_{\widehat{m}}\left(\sum_{v=1}^{V}P_{\widehat{m}, v}-1\right),
\end{align}
where $\kappa_{\widehat{m}}$ is the Lagrange multiplier. The partial derivatives of the Lagrange function w.r.t. $P_{\widehat{m}, v}$ and $g_m$ are derived as
\begin{equation}
\left\{\begin{aligned}
\frac{\partial\mathcal{L}}{P_{\widehat{m}, v}}&=-\frac{1}{A}\sum_{a=1}^{A_\mathrm{elite}}\frac{\mathbbm{1}_{\{\mathcal{I}_{\widehat{m}}(\boldsymbol{\phi}_{[a]})=\varphi_v\}}}{P_{\widehat{m}, v}}+\kappa_{\widehat{m}},\\
\frac{\partial\mathcal{L}}{g_m}&=-\frac{1}{A}\sum_{a=1}^{A_\mathrm{elite}}\frac{\mathcal{I}_{m}(\boldsymbol{\xi}_{[a]})}{g_m}+\frac{\mathcal{I}_{m}(\boldsymbol{\xi}_{[a]})-1}{1-g_m}.
\end{aligned}\right.
\end{equation}
Setting $\frac{\partial\mathcal{L}}{P_{\widehat{m}, v}}=0$ and $\frac{\partial\mathcal{L}}{g_m}=0$, we have
\begin{equation}
\left\{\begin{aligned}
P_{\widehat{m}, v}&=\frac{\sum_{a=1}^{A_\mathrm{elite}}\mathbbm{1}_{\{\mathcal{I}_{\widehat{m}}(\boldsymbol{\phi}_{[a]})=\varphi_v\}}}{A_\mathrm{elite}},~ \widehat{m}\in\widehat{M}, v\in\mathcal{V},\\
g_m&=\frac{\sum_{a=1}^{A_\mathrm{elite}}\mathcal{I}_{m}(\boldsymbol{\xi}_{[a]})}{A_\mathrm{elite}},~ m\in\mathcal{M}.
\end{aligned}\right.
\end{equation}
Note that the result of $P_{\widehat{m}, v}$ is obtained by deriving the Lagrange multiplier as $\kappa_{\widehat{m}}=\frac{A_\mathrm{elite}}{A}$ using the equality constraint in \eqref{eq_entropy_1}. It is worth noting that the obtained optimal solutions cannot be directly utilized to update the tilting parameters such that the risk of converging towards local optimal solutions is mitigated. Here, a smoothing technique is adopted to modify the tilting parameters. Specifically, the tilting parameters in the $t$-th iteration are given by
\begin{align}
P^{(t)}_{\widehat{m}, v}&\leftarrow \omega P^{(t)}_{\widehat{m}, v}+(1-\omega)P^{(t-1)}_{\widehat{m}, v},\\
g^{(t)}_m&\leftarrow \omega g^{(t)}_m+(1-\omega)g^{(t-1)}_m,
\end{align}
where $\omega\in(0, 1)$ represents the smoothing parameter, and the notation $\leftarrow$ denotes the assignment operation. We update iteratively the tilting parameters and sample the new feasible solutions until the optimal objective function values between adjacent iterations reaches a predetermined threshold.
\begin{figure}[!t]
\centering
\includegraphics[scale=0.3]{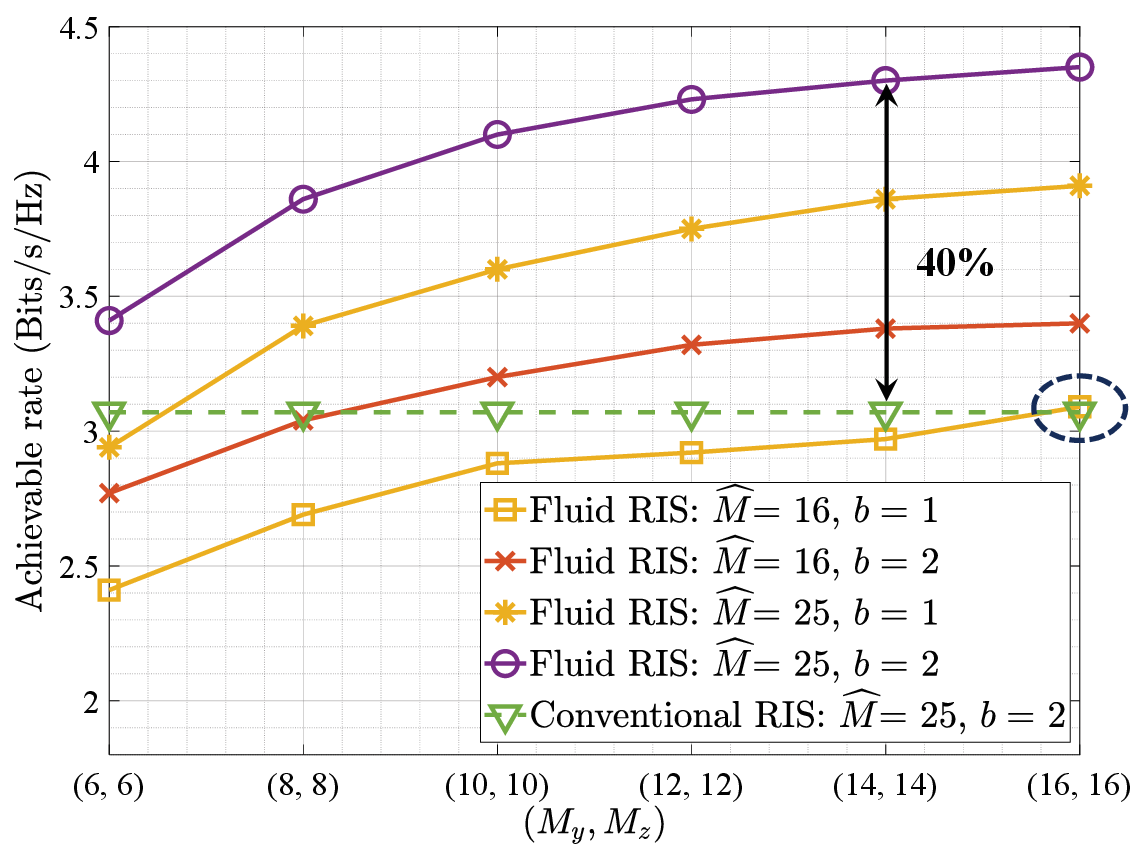}\\
\caption{The achievable rate versus the number of elements equipped at the RIS system considering different $\widehat{M}$ and phase-shift resolutions $b$.}\label{fig:ar_vs_M}
\end{figure}

\section{Simulation Results}
To evaluate the effectiveness of the proposed FRIS, we conduct a large number of simulations considering the following parameters: $P_{t\max}= 30$ dBm, $A=5(M+\widehat{M})$, $\zeta=0.05$, $\omega=0.55$, $\rho=-20$ dB, $\alpha=2.6$, $f_c=5$ GHz, $S= 2\lambda \times 2\lambda$, $d_\mathrm{br}=400$ m and $d_\mathrm{ru}=75$ m. Note that the conventional RIS scheme is utilized as the benchmark scheme, where the elements are uniformly distributed in the available region.

Initially, we examine the impact of the number of elements equipped at the RIS, $M$, on the user's achievable rate taking into account different number of activated elements and phase-shift solutions, as shown in Fig.~\ref{fig:ar_vs_M}. Specifically, it is observed that the achievable rate demonstrates a progressive increase with the increase in the number of on-off elements within the designated space across all scenarios. This can be explained by the fact that the RIS system offers a broader range of potential locations for the implementation of FRIS. As a consequence, FRIS can modulate the incident signals with more spatial DoF.

Additionally, we find that the growth in rate has a diminishing return because of the gradual reduction in potentially  valuable positions. To validate the effectiveness of FRIS, we consider the best condition ($\widehat{M}= 25, b= 2$) for conventional RIS. Simulation results indicate that under identical conditions, FRIS exhibits a significant enhancement in performance compared to conventional RIS, with this improvement becoming more pronounced as the number of available positions increases. For instance, with $M=14 \times 14 $, FRIS achieves a $40\%$ performance gain. Also, as the number of available positions increases, FRIS gradually provides superior performance gains over traditional RIS, even under less favorable implementation conditions. For example, with $ \widehat{M}=16, b=1 $, the performance of FRIS surpasses that of conventional RIS when $M=16 \times 16$. These results illustrate great potential of FRIS in enhancing the performance of wireless networks.
\begin{figure}[!t]
\centering
\includegraphics[scale=0.5]{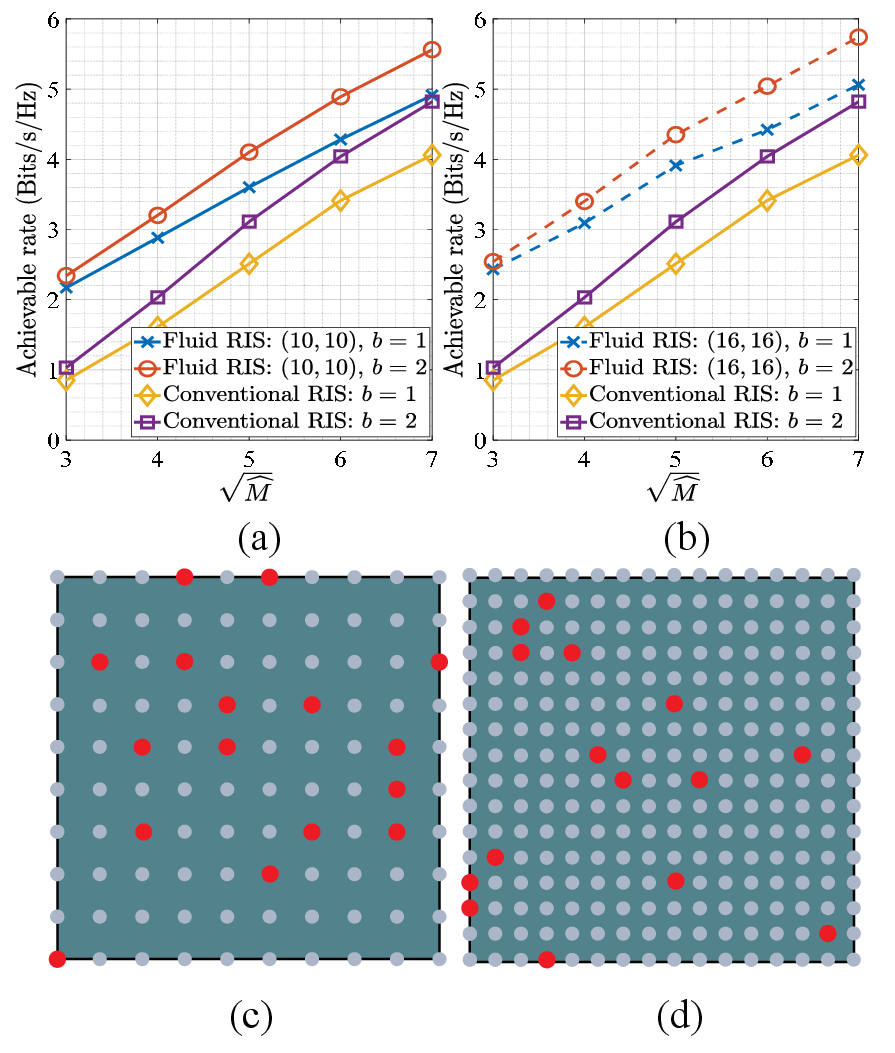}\\
\caption{The achievable rate versus $\widehat{M}$ considering different ${M}$ and $b$, as well as the optimal configuration of the FRIS under different ${M}$: (a) $(M_y, M_z)= (10, 10)$; (b) $(M_y, M_z)= (16, 16)$; (c) $(M_y, M_z)= (10, 10), \widehat{M} = 16$; (d) $(M_y, M_z)= (16, 16), \widehat{M} = 16$.}\label{fig:ar_vs_widehat_M_configurition_fluid_ris}
\end{figure}

Now we study how the achievable rate varies against the number of selected elements in FRIS, i.e., $\widehat{M}$, taking into account different RIS configurations and the number of bits for phase-shift, as shown in Fig.~\ref{fig:ar_vs_widehat_M_configurition_fluid_ris}(a) and Fig.~\ref{fig:ar_vs_widehat_M_configurition_fluid_ris}(b). In particular, we see that the achievable rate exhibits an increasing trend with the increment of the number of elements, $\widehat{M}$. This is primarily attributed to the enhanced spatial DoF provided by a greater number of elements, thereby augmenting the capability for signal modulation. Furthermore, FRIS continues to obtain substantial performance gains compared to conventional RIS, even with reduced phase shift resolution ($b =1$), with the advantage being more pronounced when $\widehat{M}$ is relatively small. This suggests that FRIS maintains a superior performance advantage under low-complexity implementation constraints. It is noteworthy that the performance gap between the FRIS and conventional RIS diminishes as $\widehat{M}$ increases. This stems from the fact that FRIS progressively converges towards the conventional RIS with increasing $\widehat{M}$. Additionally, Fig.~\ref{fig:ar_vs_widehat_M_configurition_fluid_ris}(c) and Fig.~\ref{fig:ar_vs_widehat_M_configurition_fluid_ris}(d) illustrate the optimal configurations of the FRIS with $\widehat{M}=16$ under different ${M}$, respectively. These two configurations indicate that strategically selecting the elements of the RIS system is critical for performance enhancement. It is worth noting that the precise correlation between the attainable rate and the elements' locations in the FRIS is intricate and contingent upon the instantaneous channel characteristics.

\section{Conclusion}
This letter introduced a novel FRIS framework, in which a dense matrix of on-off elements is deployed over the surface to enhance DoFs. We integrated the proposed FRIS into the SU-SISO model and formulated an optimization problem aiming to maximize the achievable rate by jointly selecting element positions and designing discrete phase shifts. We developed an iterative algorithm based on CEO to obtain a joint solution. Our simulation results demonstrated that the proposed FRIS significantly outperforms the conventional RIS counterpart.

\appendices

\ifCLASSOPTIONcaptionsoff
  \newpage
\fi
\bibliographystyle{IEEEtran}

\end{document}